# Particles redistribution and structural defects development during ice templating

Audrey Lasalle[1], Christian Guizard[1], Eric Maire[2], Jérôme Adrien[2], Sylvain Deville[1]

[1] Laboratoire de Synthèse et Fonctionnalisations des Céramiques, UMR 3080 CNRS/Saint-Gobain, 84306 Cavaillon, France

[2] Université de Lyon, INSA-Lyon, MATEIS CNRS UMR5510, F-69621 Villeurbanne, France

## Abstract

The freezing of colloidal suspensions is encountered in many natural and engineering processes. It can be harnessed through a process known as ice templating, to produce porous materials and composites exhibiting unique functional properties. The phenomenon by itself appears simple: a solidification interface propagates through a colloidal suspension. We are nevertheless still far from a complete understanding and control of the phenomenon. Such lack of control is reflected in the very large scattering of mechanical properties reported for ice-templated ceramics, largely due to the formation of structural defects. Through systematic in situ investigations, we demonstrate here the role of the suspension composition and the role of particle-particle electrostatic interactions on defect formation during ice templating. Flocculation can occur in the intercrystal space, leading to a destabilisation of the solid/liquid interface triggering the growth of crystals perpendicular to the main ice growth direction. This mechanism largely contributes to the formation of structural defects and explains, to a large extent, the scattering of compressive strength values reported in the literature.

**Keywords**
Freeze-casting, ceramic material, cellular solids, mechanical properties, defects

## Introduction

The solidification or freezing of colloidal suspensions is commonly encountered in a variety of natural processes such as the freezing of soils and the growth of sea ice. It is also seen in everyday life and engineering situations such as food engineering, cryobiology, filtration, and water purification. In materials science, the solidification of colloidal suspension is finding applications in various processes such as the processing of particle-reinforced alloys and composites, and the processing of porous materials, usually referred to as ice-templating or freeze-casting. This simple process, where a colloidal suspension is simply frozen under controlled conditions and then sublimated before sintering, provides materials with a unique porous architecture, where the porosity is almost a direct replica of the frozen solvent crystals. When a colloidal suspension is frozen unidirectionally, an initial transient regime is observed, corresponding to the initial nucleation and growth of the ice crystals. After this transient regime, a steady state regime is established, where lamellar ice crystals grow steadily along the direction imposed by the temperature gradient.

Applications of ice-templating have been demonstrated for bone substitutes [1], drug delivery [2], acoustic insulation [3], solid oxide fuel cells [4, 5] piezoelectric materials [6] and ultra-sensitive sensors [7]. The great interest in this versatile technique comes from the ease of implementation and the large range of porosity in terms of size (0.2 to 100µm), volume fraction (30 to 90%) and morphologies [8]. It was also shown that the composition of the ice-templated suspensions influences the final microstructures through the nature of additives [8] or the quantity of dispersant [9]. For any application, a proper control of the structure is of critical importance. Yet, little is understood about the dynamics of structure formation mechanisms during freezing. The characteristics of the colloidal suspension are often critical to the behaviour of the system during freezing, both in technological and natural occurrences of colloid freezing, and have rarely been analysed or understood.

A wide range of compressive strength values is reported for ice-templated materials (figure 1) when tested along their freezing direction. The compressive strength is of course dependent on composition and is greater for porous titanium or zirconia than for calcium phosphate, but the data show substantial variation even within identical systems. Because of the unusual spread in the literature data, we performed a careful review of the methods and microstructures in the literature. Microstructural observations revealed that many of the lowest strength samples in the literature had structural defects oriented perpendicular to the ice growth direction, as shown for example in figure 2b or figure 8 of reference [10]. This orientation is the worst case scenario for compressive strength measurements, and we believe that these defects are the root cause of many anomalously low strength ice templated materials found in the literature. High strength samples (figure 2a) are systematically free of such defects. The absence of such defects is clearly a necessary but not sufficient condition to obtain high compressive strength. Excessively large pore size can also lower the strength.

What we understand so far of the solidification of colloidal suspensions is derived primarily from analogies with dilute alloy systems or the investigated behaviour of single particles (or cells) in front of a moving interface. Many geological, biological, and industrial systems involve concentrated particle systems. In colloidal systems, unlike alloys, the particle-particle electrostatic interactions can strongly determine the behaviour of the system. Such aspects have not been taken into account so far. Owing to their neglect of particle-particle interactions, isolated particle models are not able to quantify the critical dependence of the final ice crystal morphologies on the initial colloid concentration – a crucially important operating parameter for industrial applications.

Through systematic in situ investigations, we demonstrate here the role of the suspension composition and the role of particle-particle electrostatic interactions on defect formation during ice templating. We performed in situ observations of crystal growth and particles redistribution by X-ray radiography and tomography. We show that particle-particle interaction can have a dramatic influence over the mechanisms controlling the formation of the structure.

| Alumina content | Nature of additive | Quantity of additive |
|---|---|---|
| $32_{vol}\%$ | $D[NH_4^+]$ | $0.2$-$0.4$-$0.7$-$1$-$2_{wt}\%$ |
| $32_{vol}\%$ | $D[NH_4^+]$ + PVA | Respectively $0.2_{wt}\%$ + $0.5_{wt}\%$ |
| $32_{vol}\%$ | $D[Na^+]$ | $0.2$-$2_{wt}\%$ |
| $32_{vol}\%$ | $d[NH_4^+]$ | $0.2$-$1_{wt}\%$ |

**Table 1.** Composition of the ice-templated alumina suspensions.

## Experimental

We developed a panel of alumina suspensions (table 1), carefully characterized by measurements of the zeta potential (figure 3a), viscosity (figure 3b), carbon organic total (COT) and observations of the state of dispersion by Cryo-FEGSEM [12]. These characterizations show that 0.2-0.4wt% of dispersant is the optimal range to obtain the strong repulsive interactions between particles necessary for an optimal dispersion state and stability, a condition traditionally required for ceramic processing routes to minimize defect formation. Adding more dispersant compresses the thickness of the diffuse layer around particles and reduces the effective range of the repulsive interactions. This causes the zeta potential values to decrease and the viscosity to increase (figure 3a).

Alumina powder (Ceralox SPA 0.5, Sasol, Tucson, AZ, USA), $D_{50} = 0.3$µm, specific surface area (SSA)= $8m^2.g^{-1}$, was dispersed in distilled water with an organic dispersant. Alumina content was held constant at 32vol%. Three sorts of suspensions were prepared, each containing a different dispersant: (1) an ammonium polymethacrylate (2) a sodium polymethacrylate and (3) an ammonium polyacrylate (respectively, Darvan CN, Darvan 7Ns, Darvan 821A Vanderbilt, Norwalk, CT, USA). These organic dispersants are respectively referred in the text as $D[NH_4^+]$, $D[Na^+]$ and $d[NH_4^+]$. With molecular weights of 13 000 and 3 500g.mol$^{-1}$ respectively, they are long organic chains of different lengths.

The dispersant concentration in each slurry was 0.2-2wt% with respect to the dried powder. Dispersant was stirred with distilled water for 30min and then the alumina powder was added. Alumina suspensions were ball-milled for 40h and de-aired before being ice-templated. In some cases, 0.5wt% (with respect to the dried powder) polyvinylic alcohol (PVA) was added as a binder.

The suspensions were first characterized by viscosity measurements performed in a concentric cylinder system (Bohlin viscosimeter, Malvern, Worcestershire, UK). The suspension was pre-sheared for 30 s followed by 30 s at rest. Viscosity was measured at a constant gradient of 50s$^{-1}$. Then a zetaprobe (Colloidal Dynamics, North Attleboro, MA, USA) was used to measure the zeta potential.

We adapted a freezing set-up on the beamline ID-19 at the European Synchrotron Radiation Facility in order to follow the freezing by X-ray radiography and subsequently observe the frozen microstructure by X-ray tomography. Suspensions were introduced into a polypropylene mold of 3mm of diameter with a syringe. Particular attention was paid to not introduce air-bubbles in suspension. The mold was placed onto a copper finger frozen from the bottom by a liquid nitrogen flux pumped from a dewar. The cooling rate was controlled by the liquid nitrogen flow rate and the temperature profile was monitored during the experiment, by a thermocouple located near the copper finger surface. The cooling rates were in the range 2-5°C.min$^{-1}$. When the cooling began, a monochromatic highly coherent X-ray beam with an energy of 20.5keV was sent through the sample. A CDD camera with 2048 x 2048 sensitive elements was placed 20mm behind the sample. The advancement of the freezing front was followed by fast acquisition radiography. For this we used a so called binning mode i.e. a reduction of the number of pixels in the projection by averaging the measurement of four neighbouring pixels from the CCD and combining them to create one pixel value. With this binning mode, we achieved a spatial resolution of 2.8µm in the radiographs. For the tomography, the frozen sample was maintained at a constant low temperature during the scan. An imaging configuration with high resolution and low acquisition speed was preferred here so the acquisition was performed without binning, with a spatial resolution of 1.4µm.

We used the sequence of radiographs to qualitatively investigate the local evolution of the concentration of colloidal particles. Alumina absorbing more than water, the X-rays absorption and thus the intensity of the signal on the radiograph is inversely related to the concentration of colloidal particles in suspension. To determine the change in colloid concentration in each image, we measured the change in intensity relative to the previous radiograph. Any increase of intensity is thus accompanied with a decrease of the local concentration of colloids (more beam coming through the suspension). Conversely, a decrease of the intensity reveals an increase of the local concentration. These variations can thus be measured qualitatively and dynamically, although not quantitatively. This image analysis was performed using the ImageJ software [13].

## Results

In situ radiography of the advancing freezing front shows a behaviour drastically dependent on the quantity of dispersant, $D[NH_4^+]$ in this case, introduced in suspension (figure 4a-c). For suspensions with an optimal dispersion (low dispersant quantity of 0.2-0.4wt%), the freezing front is composed of disoriented ice crystals above which a 430µm's thick layer of particles is accumulated. Above this accumulation, we observe a 20µm's thick particle-depleted region (figure 4a), where the particle concentration is lower than the average concentration (32vol%). Figure 4b shows that increasing the dispersant concentration to 1wt% causes the accumulated layer to decrease to 50µm. Ice crystals tend to align along the freezing direction, but the particle-depleted region remains (figure 4b).

The introduction of dispersant in large excess (2wt%) yields a cellular interface with no accumulated layer and no particle-depleted region (figure 4c). The corresponding three-dimensional tomography reconstructions of frozen microstructures show disoriented ice crystals at low quantity of dispersant (0.2wt%) (figure 4d). By increasing the dispersant quantity to 0.7wt%, ice crystals are more and more oriented (figure 4e) but defects are observed perpendicular to the freezing direction, as illustrated in figure 5 and indicated by black arrows. Since these defects are oriented perpendicular to the freezing direction, they resemble the defects found in the literature review, and shown for example in figure 2b. These defects form pores that traverse the dense ceramic walls, drastically affecting the integrity of the structure. A typical defect-free lamellar microstructure is obtained at 2wt% of $D[NH_4^+]$ (figure 4f).

This behaviour is independent of the nature of the dispersants tested here. We obtain similar microstructures with similar dispersant contents for both D[$Na^+$] and d[$NH_4^+$]. The counter ion ($NH_4^+$ or $Na^+$) and the chain length of the dispersant (3 500 to 13 000g.mol$^{-1}$) do not seem to affect the orientation of ice crystals. The same change of ice crystal morphology is observed with d[$NH_4^+$], the dispersant with a shorter chain length, with disoriented ice crystals at low quantity of dispersant (0.2wt%) (figure 6a) and a lamellar microstructure at 2wt% (figure 6b). We also obtain a similar morphological change from disoriented crystals to a lamellar microstructure with the addition of 0.5wt% of an organic binder (polyvinyl alcohol, PVA). A binder is usually required in ice-templated materials to ensure the cohesion between the particles during the freeze-drying stage, in a suspension containing a low quantity of dispersant. Thus, it is apparent that the formation of such defects is controlled largely by the quality of the dispersion, rather than other variables like dispersant counter-ion or dispersant chain length. The same morphological change is also observed when the cooling rate is increased from 2-5°C.min$^{-1}$ to 13°C.min$^{-1}$. Growth kinetics therefore also play a critical role in the mechanisms controlling the formation of the microstructure.

## Discussion

The development of defects perpendicular to the main ice growth direction during ice templating must absolutely be avoided. The microstructures obtained in conditions where such defects develop make such materials useless. We performed a review of the literature on ice growth, in particular in geophysics, and found that the perpendicular defects observed here in ice templated materials strongly resembles the ice lenses observed in geophysics. Ice lenses are ice crystals observed in frozen soils or during the directional solidification of colloidal suspension, growing perpendicular to the temperature gradient direction (for example figure 2c). A schematic view is represented in figure 2d. Ice lenses play a particularly important role in frost heave [11], by determining the soil's heave rate. The typical ice templated microstructures obtained in presence and in absence of perpendicular defects (shown in figures 2a and 2b) strongly suggest that such defects are indeed a replica of ice lenses. Since the porosity is a replica of the ice crystal network obtained after freezing during ice templating, ice lenses will result in the presence of crack-like pores perpendicular to the main ice growth direction (determined by the temperature gradient). The transverse ice crystals growing perpendicular to the main ice growth directions are therefore ice lenses.

A physically intuitive model of ice lens formation was just proposed [11], whereby the nucleation and growth of ice lens is controlled by the mechanical properties (cohesion) of the concentrated colloidal suspension between the ice crystals. The presence of heterogeneities in the concentrated suspension could therefore be a major factor facilitating the nucleation and growth of ice lenses, locally reducing the cohesion of the concentrated suspension. These heterogeneities can result from flocculation and local formation of agglomerates, and conversely particle-depleted regions. Through a systematic investigation of the panel of suspensions developed, we identified three main parameters controlling the occurrence of particle depletion and ice lenses (figure 7): the ionic strength, the viscosity of the suspension and the velocity of the interface. These three factors are experimentally controlled by the introduction of the dispersant, the quantity of additives (dispersant, binder) and the imposed cooling rate, respectively. For each of these parameters, there is a threshold value below which the behaviour of the system is changing.

The formation of particle-rich and particle-depleted regions requires a driving force for particle redistribution during freezing. Based on our results, we will first discuss the possible driving force for the formation of the accumulated particles layer, and then propose a scenario for the formation of a particle-depleted region and the nucleation and growth of ice lens, leading to the formation of structural defects.

## Origin of the particles in the layer of accumulated particles

Unidirectional ice templating usually results in the growth of lamellar ice crystals, oriented along the temperature gradient. A layer of accumulated particles above the freezing front is observed clearly in figures 4a and 4b. This seems to be related to the loss of an oriented lamellar microstructure. In addition, the presence of this layer impacts the freezing dynamics by decreasing the front velocity (figure 8). It thus plays a key role in the formation of the microstructure. The presence of regions with high and low particle concentrations implies a redistribution of the particles during freezing. The particles found in the accumulated particles layer can originate either from below, in the inter-crystals region, in which case a diffusion mechanism is probably involved, or from the non-frozen suspension above, whereby flocculation would be the driving force.

The freezing front velocities measured experimentally are in the range 5 to 30µm.s$^{-1}$. Calculations of the diffusion velocity (Appendix A1) show that the major part of particles diffuses slower than the freezing front advances. The diffusion model used considers an ideal case of spherical particles. The real diffusion coefficient is certainly lower than the one calculated here. The accumulation of particles above the tips of the ice crystals by a diffusion mechanism from below seems therefore unlikely. We propose that the accumulated particles flocculate from the depletion zone above.

## Flocculation and depletion in the intercrystal regions

To explain the occurrence of a particle-depleted region, we propose a mechanism based on the flocculation of particles induced by their progressive concentration in the intercrystal regions, driven by the ice growth. The ice crystals grow in the direction of the thermal gradient, rejecting and concentrating alumina particles surrounded by the organic dispersant (figure 9a). The system can remain stable if the velocity is high enough. In optimal condition of particle dispersion, a monolayer of dispersant is adsorbed onto particles surface by the carboxylic groups $-COO^-$ and the counter ions $NH_4^+$ or $Na^+$ are attracted by electrostatic interactions but are not bonded to particles. When the dispersant is in large excess in suspension, a monolayer is adsorbed onto particles, and the excess not adsorbed remains in suspension. The solubility of any substance in ice being extremely low ($10^{-6}$), any compound or species in solution will be rejected by the growing crystals and thus concentrated in the intercrystal regions.

When particles are concentrated between growing crystals, the non-adsorbed ionic species yield a local increase of the ionic strength. The repulsive charge layer of the alumina particles is thus compressed under ionic strength effect. The repulsive interactions are diminished and particles can locally agglomerate. Once particle agglomerates are formed, they may sediment rapidly. Calculations of the sedimentation velocity of agglomerates (Appendix A2) show that this velocity (11µm.s$^{-1}$) is compatible with the typical interface velocity (5 to 30µm.s$^{-1}$). The formation of a particle-depleted region by flocculation and sedimentation is thus compatible with the growth kinetics encountered experimentally.

## Depletion and freezing temperature

Flocculation (figures 9c) is responsible for a local increase of the freezing temperature in the particle-depleted region (figure 9d), due to particle volume fraction decrease [19], along with a decrease of the cohesive strength of the concentrated colloidal suspension. This depleted region is suddenly much more favourable to the growth of an ice lens. By the repetition of the flocculation/depletion/nucleation, more and more ice lenses are formed.

When the particle-depleted region moves above the tips of the growing ice crystals, the top of the crystals is in a zone with a higher freezing temperature. They can grow faster in this zone, resulting in an instability of the advancement of the freezing zone. Such instability has been previously attributed to the extension of a supercooling zone above the freezing front, favoured by the diffusion of particles. The formation of the supercooled zone, that we reported previously [17], therefore originates not from particles diffusion from below but by a particles flocculation from above.

For the suspensions containing a low quantity of dispersant (0.2-0.4wt%), which corresponds to suspensions with a low viscosity ($10^{-2}$Pa.s) and high zeta potential (-75mV) (figure 3), we observe that the particle-depleted region is present above the freezing front (figure 4 a,b). The displacement of the depleted region from the inter-crystals space to above the freezing front is favoured by a low viscosity and a small excess of organic dispersant in suspension.

The accumulation of particles above the freezing front and the associated disorientation of the ice crystals do not occur directly after the transient regime of freezing. We can reasonably assume that the depletion occurs first between ice crystals and then moves above the freezing front because of the progressive accumulation of particles. The conditions for this accumulation mainly depend on the dispersant quantity (which controls the ionic strength), binder addition, cooling rate and viscosity of the suspension.

When the particle-depleted region is above the freezing front, in a region free of crystals (figure 7e), nucleation and growth of new ice lenses can take place, the depleted region being in a highly supercooled state (figure 7f). These crystals grow in the depleted region, thus in a direction perpendicular to the thermal gradient. This ice lens repels and packs particles in the direction of the thermal gradient and favours the accumulation of particles above the freezing front. This leads to a repetition of the same flocculation/depletion/nucleation mechanism and the cellular morphology of the crystals is lost. This explains the disoriented microstructure observed in figure 4d.

**Parameters controlling the flocculation/depletion mechanism**

This scenario provides explanations for the effect of the various parameters identified as controlling the structuring mechanisms. The excess of dispersant or the presence of an organic binder protects the alumina particles from the agglomeration/flocculation by creating steric repulsions and it increases the cohesion of the concentrated colloidal suspension between the crystals. Moreover the presence of ionic species in large excess in aqueous media favours the depression of the freezing temperature and the curvature of the freezing interface under Gibbs-Thomson effect. This increases the supercooling degree in the area of the top of oriented ice crystals and favors higher freezing front velocity. Then if the cooling rate is increased, the freezing front velocity increases too and there is no time for flocculation to occur. In both cases, an oriented microstructure is obtained.

This also explains the previous observations of the apparent influence of particle size on the stability of the interface [17]. The previous set of experiments was performed with various powders of different granulometry, but the same mass concentration of dispersant. Decreasing the particle size with a constant dispersant mass concentration is equivalent to increasing the dispersant mass concentration with a constant particle size. For small particle sizes, all the dispersant is adsorbed onto the surface of the particles. The flocculation/depletion mechanism described here can then possibly take place, and it yields unstable interface propagation. When the particle size increases, the surface area available for adsorption decreases. The surface of the particles tends to saturate and the remaining dispersant is found in excess in the suspension, protecting the particles from the flocculation mechanism when the particle concentration increases.

## Conclusions

The flocculation/depletion mechanism exposed here can facilitate the formation of ice lenses in ice-templated materials, which are responsible to some extent of the scattering of compressive strength values. The structures with low strength correspond to processing conditions yielding a lot of ice lenses, turning into transverse cracks in the final material. The presence of the defects is of course extremely deleterious to the integrity and strength of the samples. The structures with high strength correspond to defect-free structures, obtained with a stable freezing front yielding no ice lens. Our results show that flocculation can be a viable mechanism to facilitate ice lens nucleation and growth in such system, and are a first step towards incorporating particle-particle electrostatic interactions into our understanding and modelling of the freezing of colloids.

Ice-templating is a very unusual ceramic shaping route. Contrarily to the other ceramic shaping routes, we show here that it is deleterious to optimize the dispersion state of ceramic suspension, since these conditions lead to the destabilization mechanism exposed here. Working with an excess of dispersant or binder is indeed necessary to ensure the integrity of the obtained structures.

## Acknowledgements

Financial support was provided by the National Research Agency (ANR) of France, project NACRE in the non-thematic BLANC program, reference BLAN07-2_192446. Beamline access was provided by the ERSF, under proposal MA997. Acknowledgements are due, as usual, to local staff of the beamline: Elodie Boller, Paul Tafforeau and Jean-Paul Valade for the technical and scientific support on ID-19 at ERSF. We acknowledge Jérôme Leloup, Agnès Bogner, Catherine Gauthier, Loïc Courtois and Stephen Peppin for their participation to the X-Ray experiments. Thanks to Stephen Peppin and Robert Style for providing the ice lens picture in figure 2c.

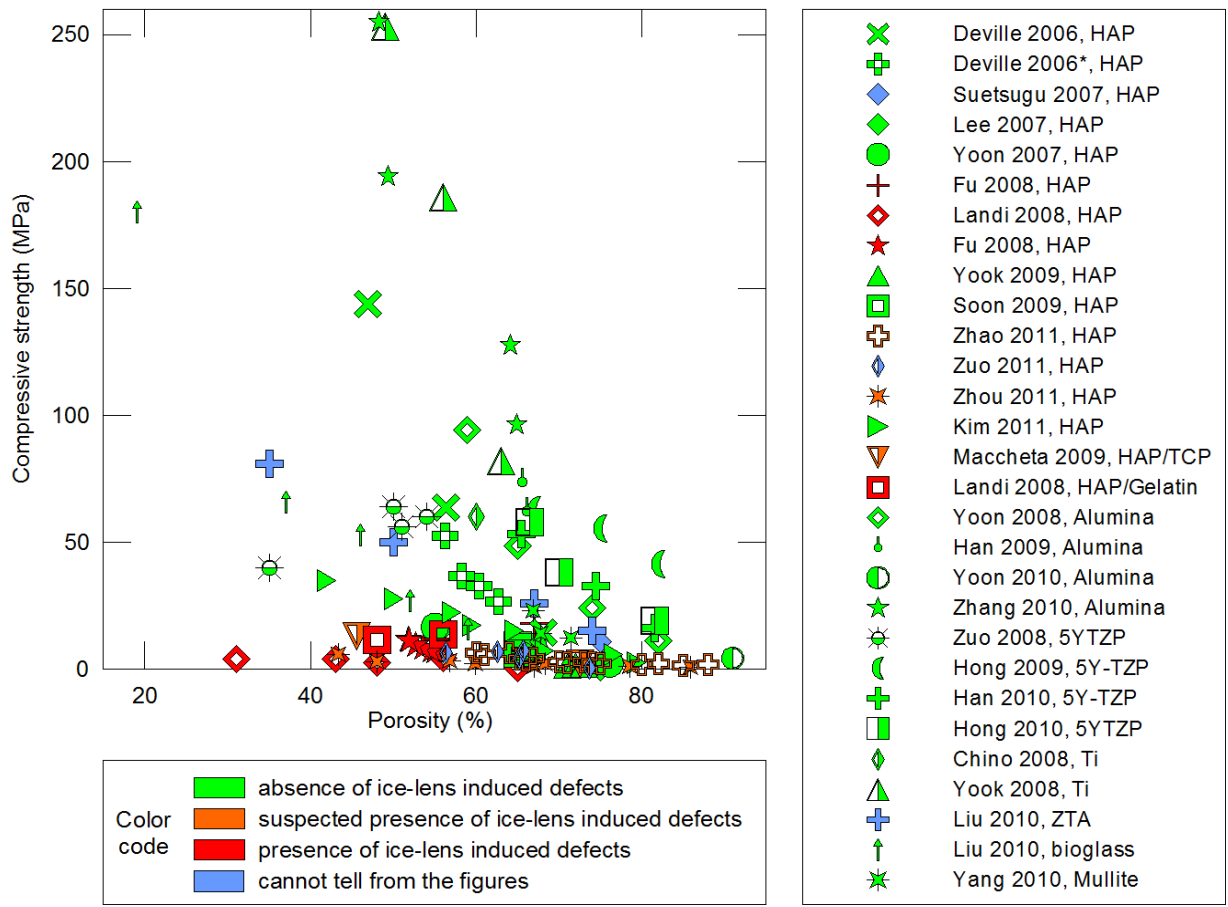

**Figure 1.** Compressive strength vs. total porosity, data from references [1, 20-46]. The colour code indicates the presence or absence of crack-like defects perpendicular to the main ice growth direction, as identified from the corresponding published figures. Such defects result from the ice lenses formation during freezing. The presence of ice-lenses induced defects is systematically correlated to a low compressive strength. A low strength can also results from excessively large pore size.

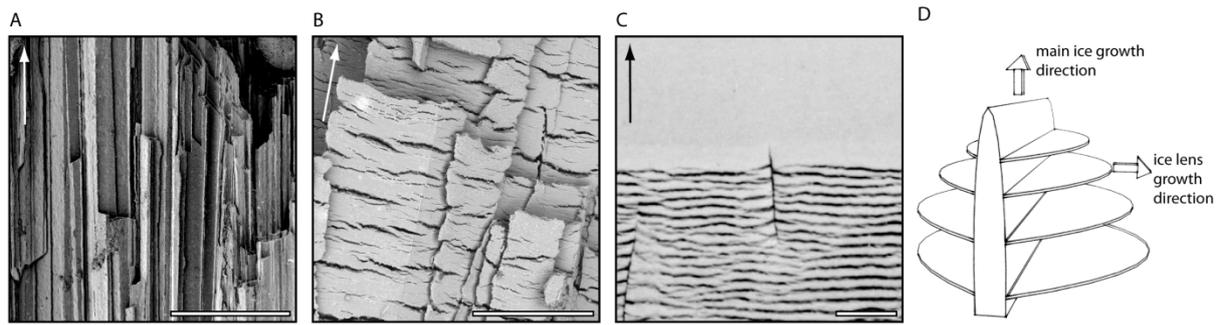

**Figure 2.** Occurrence of ice lenses and corresponding microstructures. Typical microstructure (A) without and (B) with ice-lenses induced defects, (C) ice lens in 60wt% kaolinite clay suspension, frozen unidirectionally. The concentrated kaolinite is in white; the ice lenses are the dark horizontal stripes. Arrows indicate the main ice growth direction. Ice lenses grew approximately perpendicular to the main ice growth direction. (D) Schematic representation of the typical lamellar ice crystal growing along the temperature gradient and ice lenses growing perpendicularly. Scale bars (a, b) 500µm (c) 2 mm.

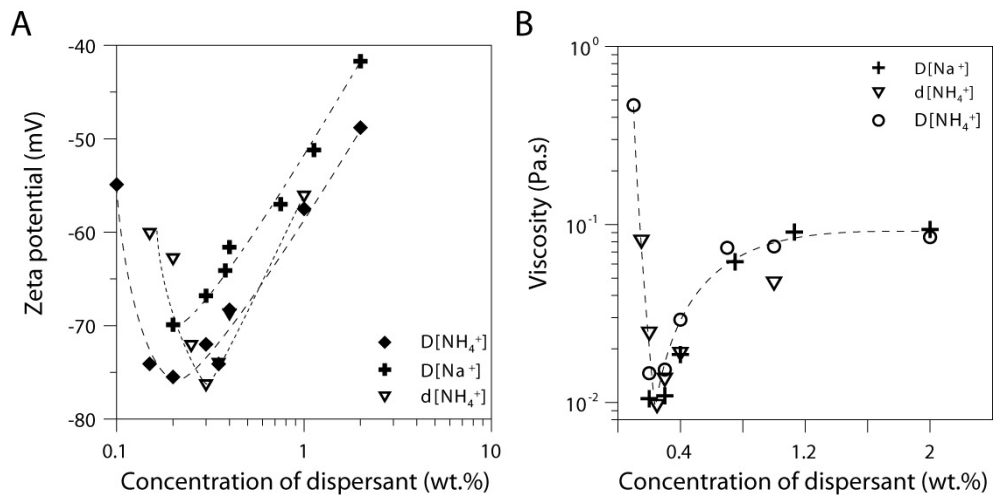

**Figure 3.** Zeta potential values [A] and viscosity values measured at 50s$^{-1}$ [B] for suspensions dispersed with D[NH$_4^+$], D[Na$^+$] or d[NH$_4^+$] with a concentration ranging from 0.2 to 2wt%.

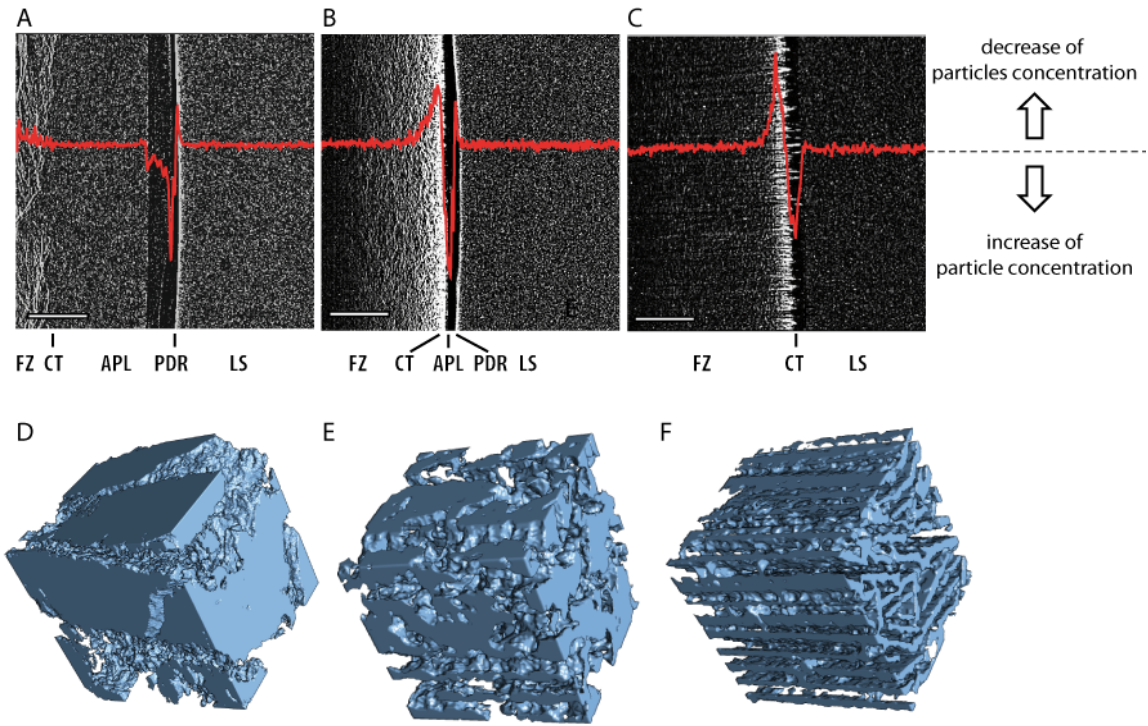

**Figure 4.** Crystal growth and particle redistribution during freezing and corresponding three-dimensional particle concentration. (A-C) In situ radiographs taken during freezing. The grey level is related to the concentration of particles. The superimposed profiles indicate the corresponding decrease or increase of particle concentration as the interface is moving from left to right, between two consecutive radiographs (see details in the experimental section). The ice-templating suspensions observed contain 0.2wt% [A], 1wt% [B] and 2wt% of D[$NH_4^+$] [C]. FZ: frozen zone, CT: crystals tips, APL: accumulated particle layer, PDR: particle-depleted region, LS: liquid suspension. Scale bar = 200µm. (D-F) 3D reconstructions from tomography of the particles rich phase regions for samples containing 0.2wt% [D], 0.7wt% [E] and 2wt% of D[$NH_4^+$] [F]. Principal ice growth direction: left to right. The black arrows show the cracks within the alumina walls, corresponding to the transverse ice crystals grown in the particle-depleted region. Scale: 140x140x140µm³.

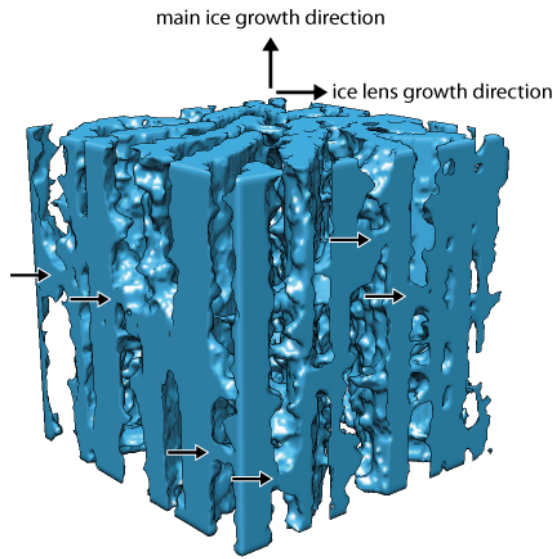

**Figure 5**: Close up view of the ice phase in presence of ice lenses. 3D reconstructions from tomography. The ice lenses are growing perpendicular to the main ice growth direction, bridging adjacent lamellar crystals. A few examples are pointed out by the arrows. Scale: 140x140x140µm³.

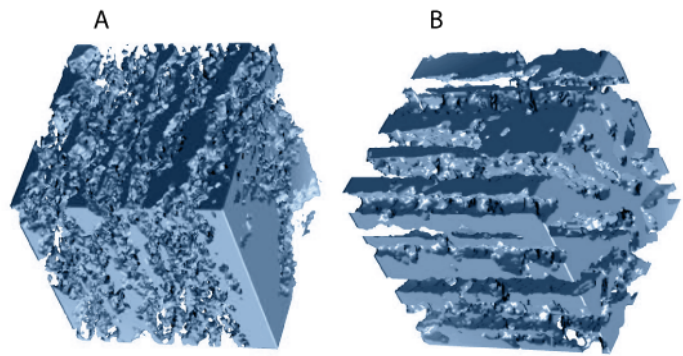

**Figure 6.** 3D reconstructions from tomography of the particle rich phase regions for samples containing 0.2wt% [A], and 1wt% [B] of d[$NH_4^+$]. Principal ice growth direction: left to right. Scale: 140x140x140µm³.

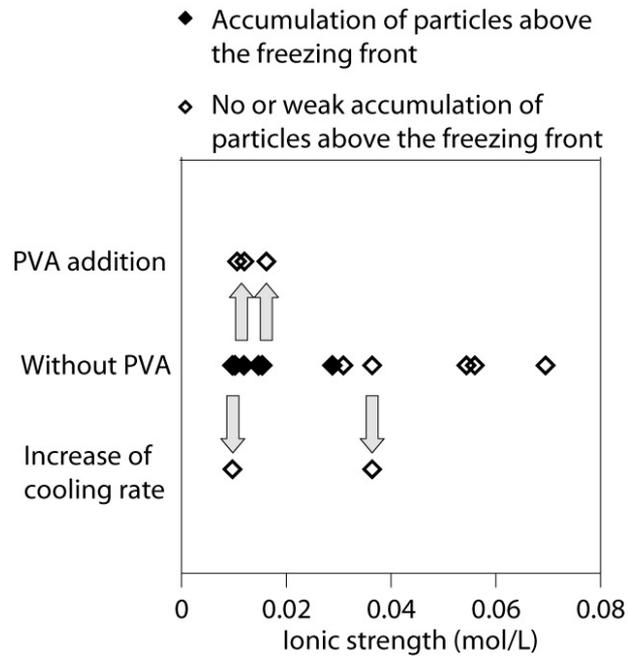

**Figure 7.** Occurrence of the accumulated particles layer as a function of ionic strength of the suspension, showing the effect of the binder addition (0.5wt% PVA) and increase of the cooling rate from 2-5°C.min$^{-1}$ to 13°C.min$^{-1}$.

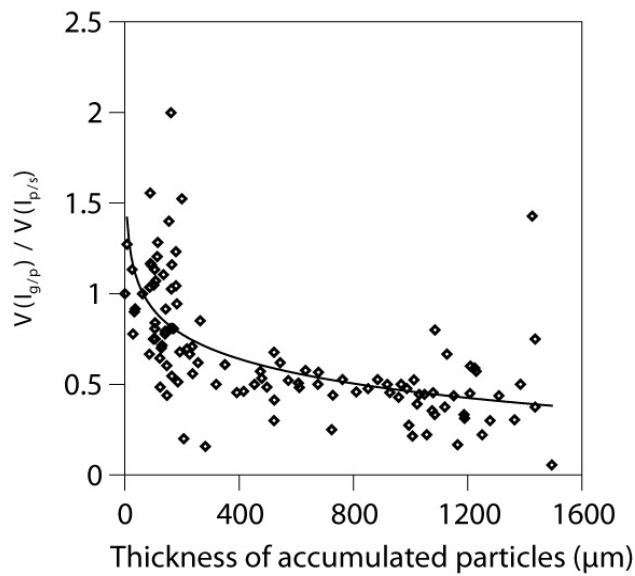

**Figure 8.** Influence of the thickness of the accumulated particles on the ice/particles interface velocity for suspensions dispersed with D[NH$_4^+$], D[Na$^+$] and d[NH$_4^+$]. All points are on the same trajectory, independently of the nature of the dispersant. The presence of an accumulated particles layer above the freezing front decreases the freezing front velocity. This decrease is more important as the thickness of the accumulated particles layer increases.

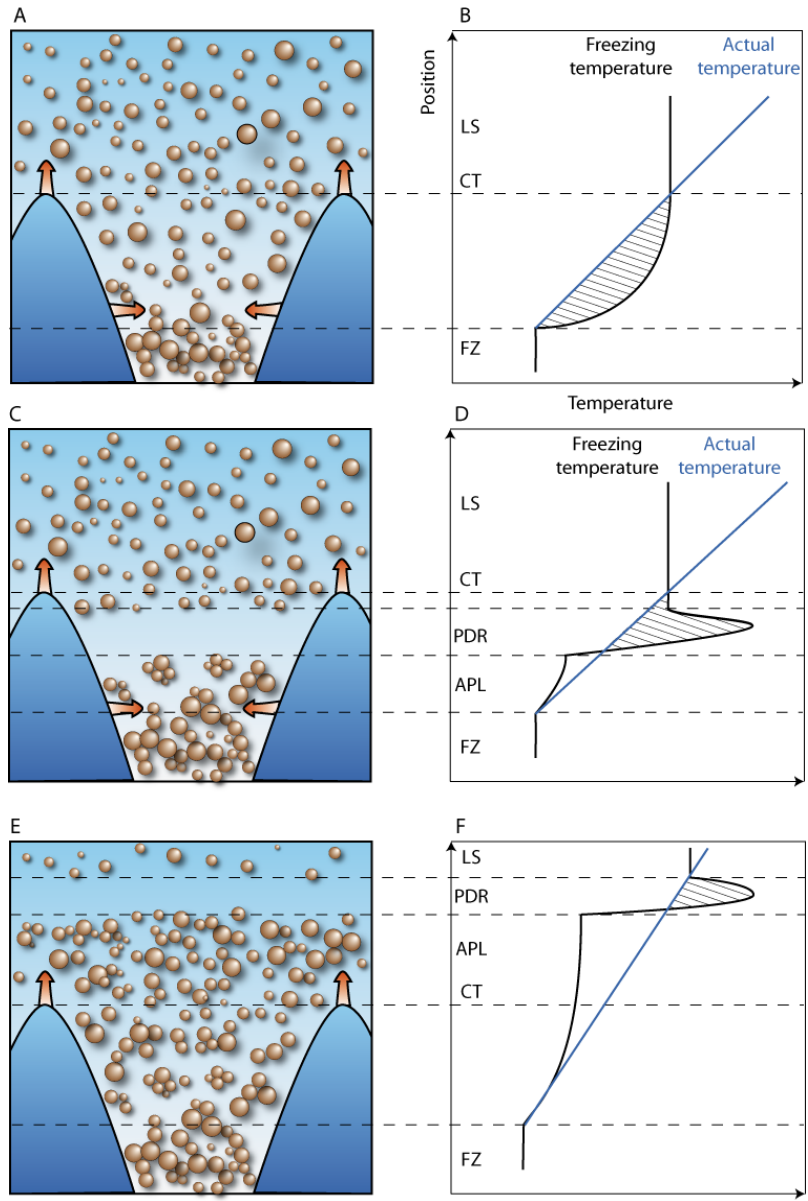

**Figure 9.** Schematic representation of the various possible situations and corresponding freezing temperature profiles. (A,B) Metastable situation with no flocculation. (C,D) Apparition of a particle-depleted region in the inter-crystals space. The growth of pre-existing crystals in the particle-depleted region is favoured by the supercooling situation. (E,F) Apparition of a particle-depleted region above the growing crystals. In absence of pre-existing crystals, nucleation and growth of an ice lens in the particle-depleted region can occur, favoured by the supercooling situation.

# Appendix

## A1. Estimation of particles diffusion velocity

We can estimate the theoretical diffusion velocity of particles by using the generic diffusion equation 1 [14].

$$L_D = 2(Dt)^{0.5} \qquad \text{(eq. 1)}$$

This equation is used to determine the distance covered by a particle during a time t with a coefficient of diffusion D. The diffusion coefficient is calculated from the equation (2) proposed by Peppin & al [15]. $D_0$ is the Stokes Einstein diffusivity and Z is the compressibility factor.

$$D(\varphi) = D_0 \widehat{D}(\varphi)$$

$$\widehat{D}(\varphi) = (1-\varphi)^6 \frac{d(\varphi Z)}{d\varphi} \qquad \text{(eq. 2)}$$

The particle size distribution (provided by the supplier) of the powder indicates that 80% of the particles are in the 50-350 nm range. By introducing the data corresponding to our system in terms of volume fraction, diameter of particles and maximal packing (obtained with the technique proposed by Liu [16] from viscosity measurements), we expect the smaller alumina particles (50nm) to diffuse at $7.5 \times 10^{-12}$ m².s⁻¹ and the larger (350 nm) at $1.2 \times 10^{-12}$ m².s⁻¹. By using equation 1, the particle diffusion velocity is estimated between 2.2 and 5.5 µm.s⁻¹.

## A2. Estimation of sedimentation velocity of agglomerates

We can estimate the sedimentation velocity, according to Wegst et al. [18]. The sedimentation velocity $v_p$ of an agglomerate of radius r is given by:

$$v_p = \frac{2(\rho_A - \rho_L)g}{9} \frac{r^2}{\eta} \qquad \text{(eq. 3)}$$

where g is the gravitational acceleration, $\eta$ the dynamic velocity of the suspension, $\rho_A$ and $\rho_L$ the density of respectively the agglomerate and the liquid. We can estimate a typical sedimentation velocity of 11µm.s⁻¹ for 2 µm agglomerates.